\documentclass[shortnote,twocolumn]{jpsj3} 
\usepackage{bm,amssymb,amsmath}	
\usepackage{graphicx}   
\usepackage{txfonts}
\newcommand{\simg}{\stackrel{>}{_\sim}}
\newcommand{\siml}{\stackrel{<}{_\sim}}

\title{
First- and second-order phase transitions between the uniform and FFLO excitonic states in the three-chain Hubbard model for Ta$_2$NiSe$_5$
}

\author{Kaoru {\sc Domon}\thanks{E-mail address: domon@phys.sc.niigata-u.ac.jp}, Takemi {\sc Yamada}, and Yoshiaki {\sc \=Ono}}

\inst{
Department of Physics, Niigata University, Ikarashi, Nishi-ku, Niigata, 950-2181, Japan 
}

\abst{
We examine the free energy and the thermodynamic properties in the three-chain Hubbard model for Ta$_2$NiSe$_5$ to clarify the phase transitions between the uniform and the FFLO excitonic states which are expected to be observed in Ta$_2$NiSe$_5$ under high pressure. 
}

\begin{document}
\maketitle

The narrow gap semiconductor Ta$_2$NiSe$_5$ shows an orthorhombic-to-monoclinic phase transition at $T_c$=328 K\cite{JLCM.116.51}, below which the flattening of the valence band top is observed in the ARPES experiments\cite{PRL.103.026402,JSNM.25.1231} and is well interpreted as excitonic condensation from a normal semiconductor to the excitonic insulator (EI) on the basis of the three-chain Hubbard model simulating a quasi-one-dimensional Ta-NiSe-Ta chain\cite{PRB.87.035121,PRB.93.041105}. The model has been investigated also for the semimetallic case\cite{JPSJ.85.053703}, where the difference in the band degeneracy between the conduction and valence bands inevitably causes the imbalance of each Fermi wavenumber and results in a remarkable Fulde-Ferrell-Larkin-Ovchinnikov (FFLO) excitonic state characterized by the condensation of excitons with finite center-of-mass momentum $q$, as expected to be observed in Ta$_2$NiSe$_5$ under high pressure\cite{pressure}. To clarify the feature of the phase transition between the uniform ($q=0$) and FFLO ($q\ne 0$) excitonic phases (EPs), this paper examines the free energy and the thermodynamic properties which have not been discussed in the previous paper\cite{JPSJ.85.053703} but would be important for the comparison with experiments under pressure\cite{pressure}. 

Our model Hamiltonian is given by
\begin{align}
H&=\sum_{k\sigma}\sum_{\alpha=1,2}\epsilon_{k}^{c}c^{\dagger}_{k\alpha\sigma}c_{k\alpha\sigma}
+\sum_{k\sigma}\epsilon_{k}^{f}f^{\dagger}_{k\sigma}f_{k\sigma}, 
\nonumber \\
 &+ V\sum_{i\alpha}\sum_{\sigma\sigma^\prime}\left(
c^{\dagger}_{i-1\alpha\sigma}c_{i-1\alpha\sigma}
+c^{\dagger}_{i\alpha\sigma}c_{i\alpha\sigma}\right)
f^{\dagger}_{i\sigma^\prime}f_{i\sigma^\prime},
\label{eq:H}
\end{align}
where $c_{k\alpha\sigma}(c_{i\alpha\sigma})$ and $f_{k\sigma}(f_{i\sigma})$ are the annihilation operators for conduction ($c$) and valence ($f$) electrons with wavenumber $k$ (site $i$), spin $\sigma=\uparrow, \downarrow$ and chain degrees of freedom for the $c$ electron $\alpha=1,2$. 
The noninteracting $c(f)$ band dispersion is given by
$
\epsilon_{k}^{c(f)}=2t_{c(f)}\left({\rm cos}k-1\right)+(-) D/2, 
$
with the hoppings $t_{c}=-0.8$ eV and $t_{f}=0.4$ eV\cite{PRB.87.035121}. $D$ is the energy gap between the $c$ and $f$ bands describing both semiconducting ($D>0$) and semimetallic ($D<0$) cases, and is varied as a decreasing function of pressure\cite{pressure}. $V$ is the $c$-$f$ Coulomb interaction crucial for the excitonic order and is set to 0.4 eV\cite{JPSJ.85.053703}.

When the condensation of excitons with center-of-mass momentum $q$ takes place, the excitonic order parameter
\[
\Delta(k,q)=-\frac{V}{N}\sum_{k^\prime}(1+e^{i(k-k^\prime)})\langle f^{\dagger}_{k^\prime+q\sigma}c_{k^\prime\alpha\sigma}\rangle
  =\Delta_{q}(1+e^{ik}e^{-i\phi_{q}})
\]
becomes finite, where $\Delta_{q}$ and $\phi_{q}$ are the magnitude and the relative phase of the order parameter, respectively\cite{JPSJ.85.053703}, and $N$ is the total number of unit cells. Within the mean-field approximation, the Hamiltonian Eq. (\ref{eq:H}) is diagonalized to yield the mean-field band dispersion as 
$
E_{k,\pm}=\epsilon_{+}(k,q)\pm\sqrt{\epsilon_{-}^{2}(k,q)+4\Delta_{q}^{2}(1+{\rm cos}(k-\phi_{q}))} 
$
with $\epsilon_{\pm}(k,q)=(\epsilon_{k}^{c}\pm\epsilon_{k+q}^{f})/2$. 
We obtain $\Delta_{q}$ and $\phi_{q}$ by solving the self-consistent equations\cite{JPSJ.85.053703}, that generally yield non-unique solutions with different values of $q$. 
Therefore, we determine the most stable solution by minimizing the free energy 
\begin{align}
F(\Delta_{q},\phi_{q}) =-\frac{k_B T}{N}\sum_{ks\sigma}{\rm ln}\left(1+e^{-(E_{ks}-\mu)/k_B T}\right)+\mu n+\frac{8\Delta_q^2}{V} 
\label{eq:F}
\end{align}
with respect to $q$, where $s$ is the band index and $\mu$ is the chemical potential determined so as to fix the number of electrons per unit cell to $n=n^{c}+n^{f}=2$.

\begin{figure}[t]
\begin{center}
\includegraphics[width=6.8cm]{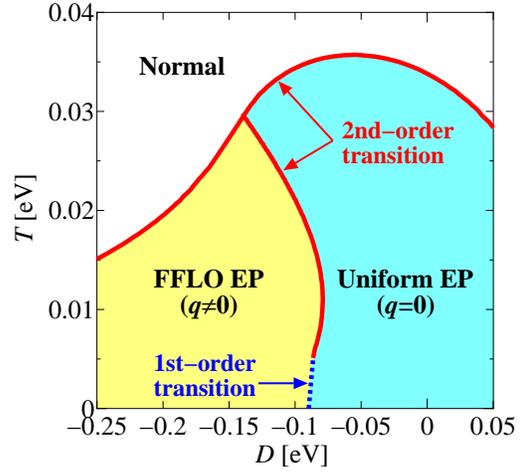}
\caption{(Color online) 
Phase diagram of the three-chain Hubbard model for Ta$_2$NiSe$_5$ as functions of the energy gap $D$ and temperature $T$ around the phase boundary between the uniform ($q=0$) and FFLO ($q\ne 0$) excitonic phases for $n$=2 and $V$=0.4 eV. Solid and dashed lines indicate the second- and first-order phase transitions, respectively. 
}
\label{Fig1}
\vspace{-0.4cm}
\end{center}
\end{figure}

Figure 1 shows the excitonic phase diagram on the $D-T$ plane around the phase boundary between the uniform and FFLO EPs. As shown in the previous paper\cite{JPSJ.85.053703}, the phase transition between the normal phase and the EPs is always second-order at the phase transition temperature $T_c$ which shows a peak around the crossover region between the BEC ($D\simg 0$) and BCS ($D\siml 0$) regimes. As for the phase transition between the uniform and FFLO EPs, the previous paper\cite{JPSJ.85.053703} has revealed that the order parameter changes continuously at high temperatures while discontinuously at low temperatures indicating the second- and first-order phase transitions, respectively (see Fig. 1). However, within numerical methods, it is difficult to exclude the possibility that the transition is continuous but very sharp. Then, we perform detailed calculations of the free energy which enables us to determine the order of the phase transition directly as shown below.

In Fig. 2, the free energy $F(\Delta_{q},\phi_{q})$ given in Eq. (\ref{eq:F}) is plotted as a function of $q$ at several values of $T$ around the phase boundary between the uniform and FFLO EPs for $D=-0.087$ eV, where a uniform-FFLO-uniform reentrant transition takes place as seen from Fig. 1. In Fig. 2(a), the minimum of $F(\Delta_{q},\phi_{q})$ shifts smoothly from zero to finite values as $T$ decreases, displaying the second-order phase transition from the uniform to FFLO EP at $T_c=0.0168$ eV. On the other hand, in Fig. 2(b), the minimum of $F(\Delta_{q},\phi_{q})$ shows a jump from a finite $q$ to zero as $T$ decreases, displaying the first-order phase transition from the FFLO to uniform EP at $T_c=0.0044$ eV. We thus confirmed the first- and second-order phase transitions shown in Fig. 1.

\begin{figure}[t]
\begin{center}
\includegraphics[width=7.2cm]{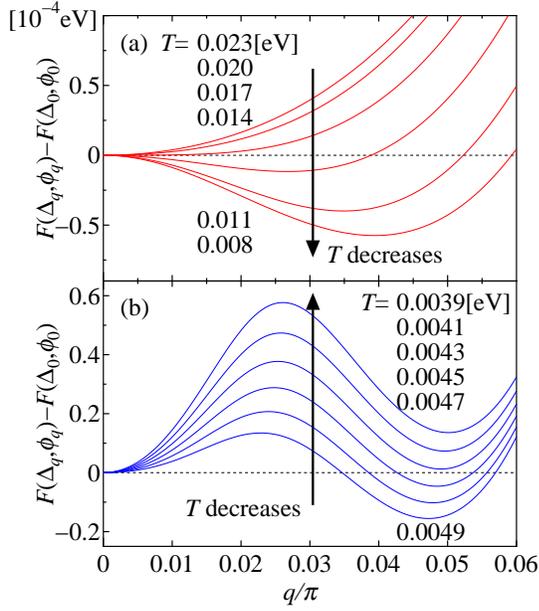}
\caption{(Color online) 
The free energy $F(\Delta_{q},\phi_{q})$ as a function of center-of-mass momentum of excitons $q$ for $D=-0.087$ eV around the phase boundary between the uniform and FFLO EPs, displaying the second-order phase transition at $T_c=0.0168$ eV (a) and the first-order one at $T_c=0.0044$ eV (b). 
}
\label{Fig2}
\vspace{-0.4cm}
\end{center}
\end{figure}

Here, we examine the thermodynamic properties in the phase boundary region. The entropy $S$ is calculated by using the following explicit form
\begin{align}
S = \frac{k_B}{N}\sum_{ks\sigma}\left\{{\rm ln}\left(1+e^{-(E_{ks}-\mu)/k_B T}\right)+\frac{E_{ks}-\mu}{k_B T} f(E_{ks})\right\}
\label{eq:S}
\end{align}
with the fermi distribution function $f(\epsilon)=[e^{(\epsilon-\mu)/k_B T} +1]^{-1}$, and then, the specific heat $C$ is obtained from the numerical derivative of $S$ with respect to $T$ or that of the internal energy 
$E = \frac{1}{N}\sum_{ks\sigma}E_{ks}f(E_{ks})+8\Delta_q^2/V$. 
Figures 3(a) and 3(b) show the $T$-dependence of $S$ and $C$ for $D=-0.087$ eV, where three phase transitions are observed as seen from the phase diagram in Fig. 1 (see also Fig. 2). As $T$ decreases, the system shows two second-order phase transitions from the normal to the uniform EP at $T_c=0.0351$ eV and from the uniform EP to the FFLO EP at $T_c=0.0168$ eV, where the jump in the specific heat for the latter transition is much smaller than that for the former one. Then, the system finally shows the first-order phase transition from the FFLO EP to the uniform EP (reentrant transition) at $T_c=0.0044$ eV with a tiny latent heat of $T\Delta S \sim 0.0044\times 0.032=0.00014$ eV per unit cell.

\begin{figure}[t]
\begin{center}
\includegraphics[width=6.5cm]{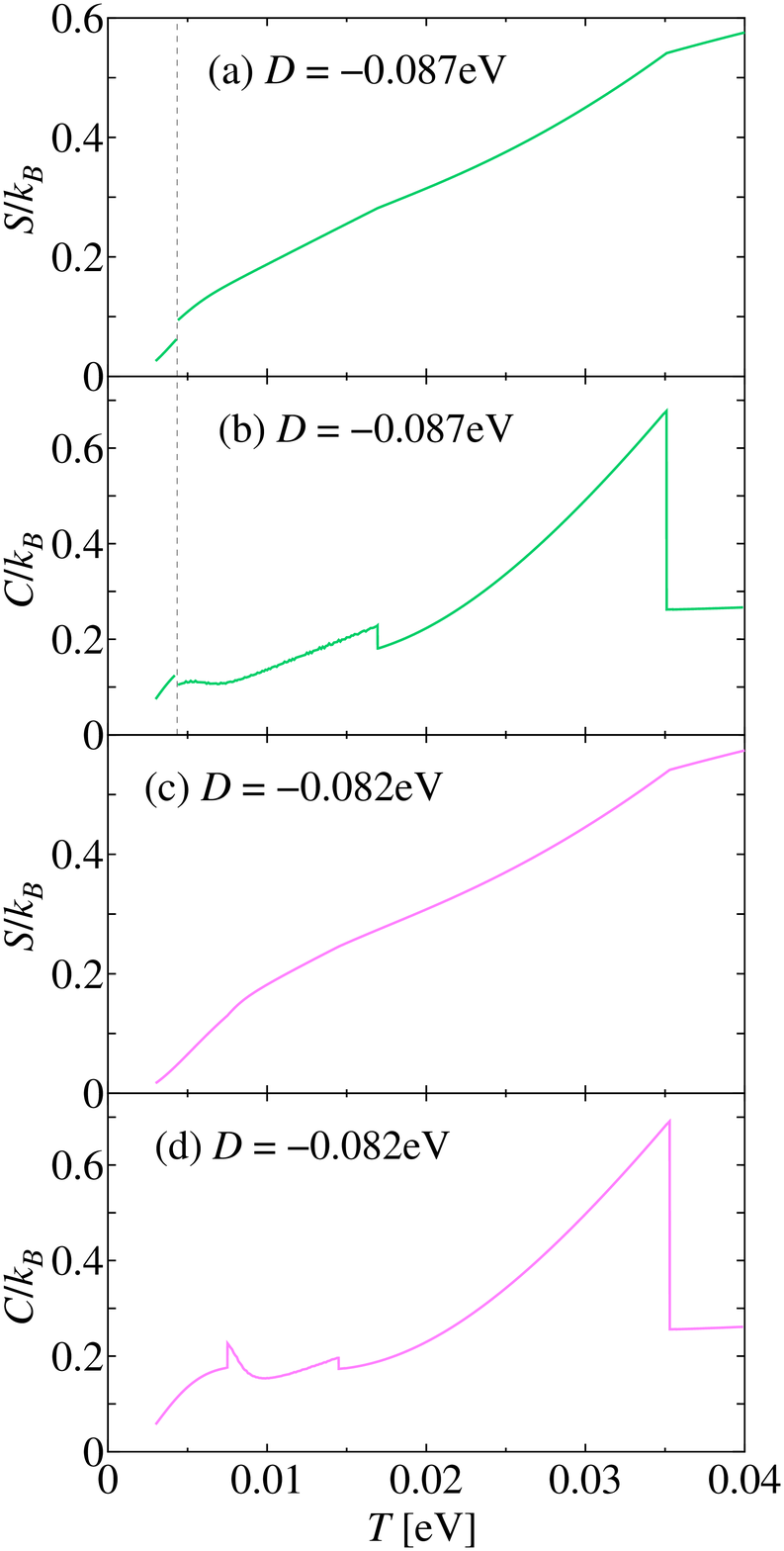}
\caption{(Color online) 
$T$-dependence of the entropy $S$ and the specific heat $C$ for $D=-0.087$ eV (a) and (b), and for $D=-0.082$ eV (c) and (d), respectively. Dotted line indicates the first-order phase transition. 
}
\label{Fig3}
\vspace{-0.4cm}
\end{center}
\end{figure}

For the case with a slightly larger value of $D=-0.082$ eV, the system shows three second-order phase transitions: normal-to-uniform EP at $T_c=0.0353$ eV, uniform-to-FFLO EP at $T_c=0.0145$ eV, and FFLO-to-uniform EP (reentrant) at $T_c=0.0075$ eV, as shown in Figs. 3(c) and 3(d). The former two transitions have typical lambda-shapes of the second-order phase transition while the third one has an anomalous mirror-writing lambda-shape which is considered to be a specific feature of the reentrant transition. Actually, such a mirror-writing lambda-shape has been observed for the magnetic-field dependence of the specific heat in the reentrant SDW phase of (TMTSF)$_2$ClO$_4$\cite{PRL.64.2054}, but has not for the temperature dependence as far as the authors know. Then, we need further investigation on the reentrant transitions including the present system from both theoretical and experimental points of view.

\begin{acknowledgments}
We would like to thank H. Fukuyama, Y. Ohta, T. Kaneko, K. Sugimoto and J. Ishizuka for valuable comments and discussions. 
This work was partially supported by a Grant-in-Aid for Scientific Research from the Ministry of Education, Culture, Sports, Science and Technology. 
\end{acknowledgments}

\bibliography{exciton_short}
\end{document}